\documentclass[10pt,conference]{IEEEtran}
\pdfoutput=1 
\usepackage{cite}
\usepackage{amsmath,amssymb}
\usepackage{amssymb}
\usepackage[pdftex]{graphicx,color}
\usepackage{psfig}
\usepackage{psfrag}
\usepackage{subfigure}
\usepackage{url}
\usepackage{array}
\usepackage{fontenc,times}
\usepackage{shadow,rotate}
\makeatletter

\let\NAT@parse\undefined
\makeatother
\ifx\Localpdfoutput\undefined
\usepackage[hypertex]{hyperref}
\else
\usepackage[pdftex,hypertexnames=false]{hyperref}
\fi
\newtheorem{theorem}{\textbf{Theorem}}
\newtheorem{corollary}{\textbf{Corollary}}
\newtheorem{proposition}{\textbf{Proposition}}

\newtheorem{remark}{Remark}
\newcommand{\mc}{\mathcal} 

\begin{document}
{\fontencoding{OT1}\fontsize{9.4}{11.25pt}\selectfont
\bibliographystyle{IEEEtran}
\title{Cooperative Relaying with State Available at the Relay}

\author{Abdellatif Zaidi$\:^{\dagger}$ \qquad Shivaprasad Kotagiri$\:^{\ddagger}$ \qquad J. Nicholas Laneman$\:^{\ddagger}$ \qquad Luc Vandendorpe$\:^{\dagger}$\vspace{0.3cm}\\
$^{\dagger}$ \'Ecole Polytechnique, Universit\'{e} Catholique de Louvain, LLN-1348, Belgium\\
$^{\ddagger}$ Department of Electrical Engineering, University of Notre, Notre Dame, IN-46556, USA\\
\{zaidi,vandendorpe\}@uclouvain.be, \{skotagir,jnl\}@nd.edu\\
\thanks{The work by A. Zaidi and L. Vandendorpe was supported in part by the EU framework program COOPCOM and the Walloon region framework program NANOTIC-COSMOS. The work by S. Kotagiri and J. Nicholas Laneman was supported in part by the State of Indiana through the Twenty-First Century Research and Technology Fund.}
}
\maketitle

\begin{abstract}
We consider a state-dependent full-duplex relay channel with the state of the channel non-causally available at only the relay. In the framework of cooperative wireless networks, some specific terminals can be equipped with cognition capabilities, i.e, the relay in our model. In the discrete memoryless (DM) case, we derive lower and upper bounds on channel capacity. The lower bound is obtained by a coding scheme at the relay that consists in a combination of codeword splitting, Gel'fand-Pinsker binning, and a decode-and-forward scheme. The upper bound is better than that obtained by assuming that the channel state is available at the source and the destination as well. For the Gaussian case, we also derive lower and upper bounds on channel capacity. The lower bound is obtained by a coding scheme which is based on a combination of codeword splitting and Generalized dirty paper coding. The upper bound is also better than that obtained by assuming that the channel state is available at the source, the relay, and the destination. The two bounds meet, and so give the capacity, in some special cases for the degraded Gaussian case.
\end{abstract}

\section{Introduction}\label{secI}
Channels that depend on random parameters have received considerable attention over the last decade, due to a wide range of possible applications. This includes single user models \cite{Sh58,GP80,C83,HG83} and multiuser models (see, e.g., \cite{GP84,S05,ZSE02} and references therein). For multiuser models, one key issue in the study of state-dependent channels is whether the parameters controlling the channel are known {\it symmetrically}, i.e., to all, or {\it asymmetrically}, i.e., to only some of, the users in the communication model. The broadcast channel (BC) with state available at the transmitter but not at the receivers is considered in \cite{GP84,S05,SS05}. The multiple access channel (MAC) with partial state information at all the encoders and full state information at the decoder is considered in \cite{CS05}.

In the Gaussian case, the MAC with all informed encoders, the BC with informed encoder, the physically degraded relay channel (RC) with informed source and informed relay, and the physically degraded relay broadcast channel (RBC) with informed source and informed relay are studied in \cite{GP84,KSS04,ZV07b}. In all these cases, it is shown that some variants of Costa's dirty paper coding (DPC) \cite{C83} achieve the respective capacity or the respective capacity region. Also, since for all these models the variant of DPC achieves the trivial upper or outer bound obtained by assuming that the channel state is also available at the decoders in the model, it is not required to obtain any non-trivial upper or outer bounds. For all these models, the key assumption that makes the problem relatively easy is the availability of the channel state at {\it all} the encoders in the communication model. It is interesting to study state-dependent multi-user models in which {\it only some}, i.e., not all, the encoders are informed about the channel state, because the uninformed encoders in the model cannot apply DPC.

The state-dependent MAC with some, but not all, encoders informed of the channel state is considered in \cite{KL04,KL07a,KL07,SBSV07,SBSV07a} and the state-dependent relay channel with only informed source is considered in \cite{ZV07b,ZV08d}. For all these models, in the Gaussian case, the informed encoder applies a slightly generalized DPC (GDPC) in which the channel input and the channel state are correlated. In these models, the uninformed encoders benefit from the GDPC applied by the informed encoders because the negative correlation between the codewords at the informed encoders and the channel state can be interpreted as partial state cancellation. For the state-dependent MAC with one informed encoder, the capacity region for the Gaussian case is obtained by deriving a non-trivial upper bound in the case in which the message sets are degraded \cite{SBSV07}. 

In this work, we consider a three terminal state-dependent full-duplex RC in which the state of the channel is known non-causally to only the relay, i.e., but neither to the source nor to the destination. We refer to this communication model as state-dependent RC with informed relay. This model is shown in Figure~\ref{StateDependentDiscreteMemorylessRelayChannel}. 
\begin{figure}[htpb]
\vspace{-0.3cm}
\centering
\resizebox{0.9\linewidth}{!}{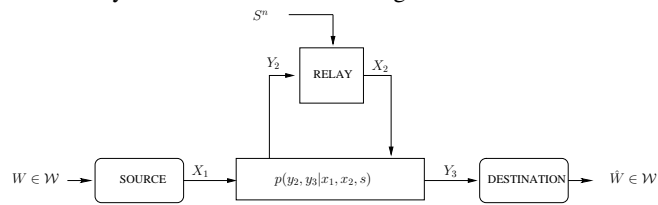}
\caption{RC with state information available non-causally at only the relay.}
 \label{StateDependentDiscreteMemorylessRelayChannel}
\vspace{-0.3cm}
 \end{figure}

For the discrete memoryless (DM) case, we derive a lower bound (Section \ref{secIII}) and an upper bound (Section \ref{secIV}) on the capacity of the state-dependent general, i.e., not necessarily degraded, RC with informed relay. The lower bound is obtained by a coding scheme at the relay that uses a combination of codeword splitting, Gel'fand-Pinsker coding, and regular encoding backward decoding \cite{C82} for full-duplex decode-and-forward (DF) relaying \cite{CG79}. The upper bound on the capacity is better than that obtained by assuming that the channel state is also available at the source and the destination. Also, this upper bound is non-trivial and connects with a bounding technique which is developed in the context of multiple access channels with asymmetric channel state in \cite[Theorem 2]{SBSV07a}. However, we note that, in this paper, the upper bound is proved using techniques that are different from those in \cite{SBSV07a}.  Furthermore, we also specialize the results in the DM case to the case in which the channel is physically degraded.

For the Gaussian case (Section \ref{secV}), we derive lower and upper bounds on channel capacity by applying the concepts developed in the DM case to the case in which the CSI is additive Gaussian, i.e., models an additive Gaussian interference, and the ambient noise is additive Gaussian. Furthermore, we point out the loss caused by the asymmetry and show that the lower bound is (in general) close and is tight in a number of special cases if the channel is physically degraded. The key idea for the lower bound is an appropriate code construction which allows the source and the relay to send coherent signals (by enabling correlation between source and relay signals, though only one of the two encoders is informed) and, at the same time, have the source possibly benefit from the availability of the CSI at the relay (through a generalized DPC). Also, we characterize the optimal strategy of the relay balancing the trade-off between enhancing source's transmitted signal and combating the interference. Section \ref{secVI} presents some illustrative numerical examples. 
 
\section{The DM RC With Informed Relay Only}\label{secII}

The state-dependent DM RC, denoted by $({\mc X_1}\times{\mc X_2},W_{Y_2,Y_3|X_1,X_2,S},{\mc Y_2}\times{\mc Y_3},\mc S)$ and depicted in Figure~\ref{StateDependentDiscreteMemorylessRelayChannel}, consists of five finite sets $\mc X_1, \mc X_2, \mc S, \mc Y_2, \mc Y_3$, a distribution $Q_S$ on the set $\mc S$ and a family of conditional probability distributions $W_{Y_2,Y_3|X_1,X_2,S}$ from ${\mc X_1}\times{\mc X_2}\times{\mc S}$ to ${\mc Y_2}\times{\mc Y_3}$. Let $X_1^n=(X_{1,1},\cdots,X_{1,n})$ and $X_2^n=(X_{2,1},\cdots,X_{2,n})$ designate the inputs of the source and the relay respectively, and $Y_2^n=(Y_{2,1},\cdots,Y_{2,n})$ and $Y_3^n=(Y_{3,1},\cdots,Y_{3,n})$ designate the outputs of the relay and the channel, respectively. We assume that the channel states $S^n$ are i.i.d., each distributed according to $Q_S$. 

As it can be seen from Figure~\ref{StateDependentDiscreteMemorylessRelayChannel}, the setup we consider is asymmetric in the sense that only the relay is informed of the channel states. The relay observes the CSI non-causally, and we refer to this channel as state-dependent RC with informed relay. Also, the channel will be said to be physically degraded if the conditional distribution $W_{Y_2,Y_3|X_1,X_2,S}$ can be written as
\begin{equation}
W_{Y_2,Y_3|X_1,X_2,S}=W_{Y_2|X_1,X_2,S}W_{Y_3|Y_2,X_2,S}.
\label{DistributionDegradedChannel}
\end{equation}

In Section \ref{secIII} and Section \ref{secIV} we establish bounds on the capacity of the state-dependent RC with informed relay in the DM case. We assume that the alphabets $\mc S, \mc X_1, \mc X_2$ are finite. The proofs of these bounds are rather lengthy and, for the sake of brevity, they are either outlined only or omitted here. Detailed proofs are reported in \cite{ZKLV08b}.

\section{DM Case: Lower Bound on Capacity}\label{secIII}
The following theorem provides a lower bound on the capacity of the state-dependent DM RC with informed relay.
\vspace{0.2cm}

\begin{theorem}\label{TheoremAchievabeRateNonCausalCaseDiscreteMemorylessChannel}
The capacity of the state-dependent DM RC with informed relay is lower bounded by
\begin{align}
R^{\text{in}} = \max \min \Big\{&I(X_1;Y_2|S,U_1), \nonumber\\
&I(X_1,U_1,U_2;Y_3)-I(U_2;S|U_1)\Big\},
\label{AchievabeRateNonCausalCaseDiscreteMemorylessChannel}
\end{align}
where the maximization is over all joint measures $P_{S,U_1,U_2,X_1,X_2,Y_2,Y_3}$ of the form
\begin{align}
&P_{S,U_1,U_2,X_1,X_2,Y_2,Y_3}=\nonumber\\
&\hspace{0.8cm}Q_SP_{U_1}P_{X_1|U_1}P_{U_2|U_1,S}P_{X_2|U_1,U_2,S}W_{Y_2,Y_3|X_1,X_2,S}
\label{MeasureForAchievabeRateNonCausalCaseDiscreteMemorylessChannel}
\end{align}
and $U_1$ and $U_2$ are auxiliary random variables with cardinality bounded as
\begin{subequations}
\begin{align}
\label{BoundsOnCardinalityOfAuxiliaryRandonVariableU1ForAchievabeRateNonCausalCaseDiscreteMemorylessChannel}
&|\mc U_1| \leq |\mc S||\mc X_1||\mc X_2|+1\\
&|\mc U_2| \leq \Big(|\mc S||\mc X_1||\mc X_2|+1\Big)|\mc S||\mc X_1||\mc X_2|,
\label{BoundsOnCardinalityOfAuxiliaryRandonVariableU2ForAchievabeRateNonCausalCaseDiscreteMemorylessChannel}
\end{align}
\label{BoundsOnCardinalityOfAuxiliaryRandonVariablesForAchievabeRateNonCausalCaseDiscreteMemorylessChannel}
\end{subequations}
respectively.
\end{theorem}
\vspace{0.2cm}

The proof of Theorem \ref{TheoremAchievabeRateNonCausalCaseDiscreteMemorylessChannel} is based on a random coding scheme the basic idea of which is conveyed in the following remarks. 
\begin{remark}\label{remark1}
 
The lower bound \eqref{AchievabeRateNonCausalCaseDiscreteMemorylessChannel} is based on the relay operating in a decode-and-forward (DF) relaying scheme \cite[Theorem 1]{CG79}. In DF strategies, the source knows what cooperative information the relay transmits. Consequently, the source generates the codeword that achieves \textit{coherence} gains as in multi-antenna transmission, by having the channel input of the source correlated with that of the relay. In our model, at one hand, if the relay generates its channel input as a function of both the cooperative information and the channel state, then it is not possible for the source to know the relay input, because the source does not know the channel state. On the other hand, the relay should not completely ignore the known channel state which it can use to remove the effect of the CSI on the communication. To resolve the existing tension at the relay, we generate two codebooks. In one codebook, the random codewords $U_1^n$ are generated using a random variable $U_1$ which is independent of the channel state $S$. The relay chooses the appropriate random codeword from this codebook using the cooperative information only. In another codebook, the codewords $U_2^n$ are generated using the random variable $U_2$ which is correlated with the channel state $S$ and $U_1$ through $P_{U_2|U_1,S}$. The relay chooses the codeword from this codebook using both the cooperative information and the channel state, to remove the effect of the channel state on the communication. Then the relay generates the channel input $X_2^n$ from $(U_1^n,U_2^n)$ using the conditional probability law $P_{X_2|U_1,U_2}$. Likewise, the source knows $U_1^n$ because it is a function of only the cooperative information, and, given $U_1^n$, it generates the random codeword $X_1^n$ according to the conditional probability law $P_{X_1|U_1}$. Thus the channel inputs of the source and the relay are correlated through $U_1^n$. This is the main idea of the coding scheme that we use at the relay.

\end{remark}
\begin{remark}
The term $[I(X_1,U_1,U_2;Y_3)-I(U_2;S|U_1)]$ in  \eqref{AchievabeRateNonCausalCaseDiscreteMemorylessChannel} illustrates the multi-antenna system behavior when the encoders are asymmetrically informed about the channel: it can be interpreted as an achievable sum rate over a two-encoder MAC with only one encoder being informed of the CSI. In \cite{SBSV07,KL07}, the authors derive the capacity region of a state-dependent MAC with one informed encoder in the case in which the informed encoder knows the message to be transmitted by the uninformed encoder. In our model, transmission from the source and  the relay to the destination can also be viewed as that over a MAC with one informed encoder. However, in our case, due to the structure of the DF scheme, the situation is different from \cite{SBSV07,KL07} since the uninformed encoder (the source) knows the message of the informed encoder (the relay), not the opposite. This makes coding at the relay more difficult in our setup.
\end{remark}

\section{DM Case: Upper Bound on Capacity}\label{secIV}
The following theorem provides an upper bound on the capacity of the state-dependent DM RC with informed relay.
\vspace{0.2cm} 

\begin{theorem}\label{TheoremOuterBoundNonCausalCaseDiscreteMemorylessChannel}
The capacity of the state-dependent DM RC with informed relay is upper bounded by
\begin{align}
R^{\text{out}} \: =\:\: \max \min \Big\{&I(X_1;Y_2,Y_3|S,X_2), \nonumber\\
&I(X_1,X_2;Y_3|S)-I(X_1;S|Y_3)\Big\},
\label{OuterBoundNonCausalCaseDiscreteMemorylessChannel}
\end{align}
where the maximization is over all joint measures $P_{S,X_1,X_2,Y_2,Y_3}$ of the form
\begin{align}
&P_{S,X_1,,X_2,Y_2,Y_3}=Q_SP_{X_1}P_{X_2|X_1,S}W_{Y_2,Y_3|X_1,X_2,S}.
\label{MeasureForOuterBoundNonCausalCaseDiscreteMemorylessChannel}
\end{align}
\end{theorem}
\vspace{0.2cm}

In the second term of the minimization in \eqref{OuterBoundNonCausalCaseDiscreteMemorylessChannel}, the term $I(X_1;S|Y_3)$ can be interpreted as the rate penalty in the information conveyed to the destination caused by not knowing the channel state at the source as well. This rate loss makes the above upper bound tighter than the trivial upper bound which is obtained by assuming that the channel state is also available at the source and the destination, i.e.,
\begin{align}
&R^{\text{out}}_{\text{triv}} \: = \:\: \max \min \big\{I(X_1;Y_2,Y_3|S,X_2),I(X_1,X_2;Y_3|S)\big\},
\label{TrivialOuterBoundNonCausalCaseDiscreteMemorylessChannel}
\end{align}
where here maximization is over all joint measures $P_{S,X_1,X_2,Y_2,Y_3}$ of the form
\begin{align}
&P_{S,X_1,,X_2,Y_2,Y_3}=Q_SP_{X_1|S}P_{X_2|X_1,S}W_{Y_2,Y_3|X_1,X_2,S}.
\label{MeasureForTrivialOuterBoundNonCausalCaseDiscreteMemorylessChannel}
\end{align}
If the channel is physically degraded, the upper bound in Theorem \ref{TheoremOuterBoundNonCausalCaseDiscreteMemorylessChannel} reduces to the one in the following corollary.
\begin{corollary}\label{CorollaryOuterBoundDegradedRelayChannelNonCausalCaseDiscreteMemorylessChannel}
The capacity of the state-dependent physically degraded DM RC with informed relay is upper bounded by
\begin{align}
R^{\text{out}}_{\text{D}} \:=\:\: \max \min \Big\{&I(X_1;Y_2|S,X_2), \nonumber\\
&I(X_1,X_2;Y_3|S)-I(X_1;S|Y_3)\Big\}
\label{OuterBoundDegradedChannelNonCausalCaseDiscreteMemorylessChannel}
\end{align}
where the maximization is over all probability distributions of the form
\begin{align}
P_{S,X_1,X_2,Y_2,Y_3}&=Q_SP_{X_1}P_{X_2|X_1,S}W_{Y_2|X_1,X_2,S}W_{Y_3|Y_2,X_2,S}.
\label{MeasureForOuterBoundDegradedChannelNonCausalCaseDiscreteMemorylessChannel}
\end{align}
\end{corollary}
\section{The Gaussian Relay Channel}\label{secV}
In this section, we consider a state-dependent full-duplex Gaussian RC in which both the channel state and the noise are additive and Gaussian. In this model also, we assume that the additive channel state is non-causally known to only the relay. 
\subsection{Channel Model}\label{secV_subsecA}
For our model of the state-dependent full-duplex Gaussian RC with informed relay, the channel outputs $Y_{2,i}$ and $Y_{3,i}$ at time instant $i$ for the relay and the destination, respectively, are related to the channel input $X_{1,i}$ from the source and $X_{2,i}$ from the relay, and the channel state $S_i$ by
\begin{subequations}
\begin{align}
\label{ReceivedAtRelayFullDuplexRegimeGaussianRCWithState}
Y_{2,i}&=X_{1,i}+S_i+Z_{2,i},\\
Y_{3,i}&=X_{1,i}+X_{2,i}+S_i+Z_{3,i},
\label{ReceivedAtDestinationFullDuplexRegimeGaussianRCWithState}
\end{align}
\label{ChannelModelForFullDuplexRegimeGaussianRCWithState}
\end{subequations}
where $S_i$ is a zero mean Gaussian random variable with variance $Q$, $Z_{2,i}$ is zero mean Gaussian with variance $N_2$, and $Z_{3,i}$ is zero mean Gaussian with variance $N_3$. The random variables $S_i$, $Z_{2,i}$ and $Z_{3,i}$ at time instant $i \in \{1,2,\ldots,n\}$ are mutually independent,  are independent of the channel inputs $(X_1^n,X_2^n)$ , and are independent of $(S_j, Z_{2,j}, Z_{3,j})$ for $j \neq i$.\\

For the full-duplex degraded additive Gaussian RC, the channel outputs $Y_{2,i}$ and $Y_{3,i}$ for the relay and the destination, respectively, are related to the channel inputs $X_{1,i}$ and $X_{2,i}$ and the state $S_i$ by
\begin{subequations}
\begin{align}
\label{ReceivedAtRelayFullDuplexRegimeDegradedGaussianRCWithState}
Y_{2,i}&=X_{1,i}+S_i+Z_{2,i},\\
Y_{3,i}&=X_{2,i}+Y_{2,i}+Z'_{3,i},
\label{ReceivedAtDestinationFullDuplexRegimeDegradedGaussianRCWithState}
\end{align}
\label{ChannelModelForFullDuplexRegimeDegradedGaussianRCWithState}
\end{subequations}
where $(Z'_{3,1},\cdots,Z'_{3,n})$ is a sequence of i.i.d. zero mean Gaussian random variables with variance $N'_3=N_3-N_2$ which is independent of $Z^n_2$.

We consider individual power constraints on the transmitted power, $\sum_{i=1}^{n}X_{1,i}^2 \leq nP_1, \:\: \sum_{i=1}^{n}X_{2,i}^2 \leq nP_2$. 

\subsection{Bounds on Capacity}\label{secV_subsecB}
The results obtained in Section \ref{secIV} for the DM case can be applied to memoryless channels with discrete time and continuous alphabets using standard techniques \cite{G68}. We use the bounds in Theorem \ref{TheoremAchievabeRateNonCausalCaseDiscreteMemorylessChannel} and Theorem \ref{TheoremOuterBoundNonCausalCaseDiscreteMemorylessChannel} to compute bounds on channel capacity for the Gaussian case.

The following theorem provides a lower bound on the capacity of the state-dependent Gaussian RC with informed relay.
\vspace{0.2cm}

\begin{theorem}\label{TheoremAchievabeRateNonCausalCaseGaussianChannelFullDuplexRegime}
The capacity of the state-dependent general Gaussian RC with informed relay is lower bounded by
\begin{align}
& R^{\text{in}}_{\text{G}}=\max_{\rho'_{12}} \min \Big\{\frac{1}{2}\log(1+\frac{P_1(1-\rho'^2_{12})}{N_2}),\nonumber\\
&\hspace{1.5cm}\max_{\theta, \rho'_{2s}}\:\:\frac{1}{2}\log\Big(1+\frac{P_1+\bar{\theta}P_2+2\rho'_{12}\sqrt{\bar{\theta}P_1P_2}}{{\theta}P_2+Q+N_3+2\rho'_{2s}\sqrt{{\theta}P_2Q}}\Big)\nonumber\\
&\hspace{3.5cm}+\frac{1}{2}\log(1+\frac{{\theta}P_2(1-\rho'^2_{2s})}{N_3})\Big\},
\label{AchievabeRateNonCausalCaseGaussianChannelFullDuplexRegime}
\end{align}
where the maximization is over parameters $\rho'_{12} \in [0,1]$, $\theta \in [0,1]$, $\rho'_{2s} \in [-1,0]$, and $\bar{\theta}=1-\theta$.
\end{theorem}
\vspace{0.2cm}

The proof of Theorem \ref{TheoremAchievabeRateNonCausalCaseGaussianChannelFullDuplexRegime} is based on the evaluation of the lower bound \eqref{AchievabeRateNonCausalCaseDiscreteMemorylessChannel} with an appropriate jointly Gaussian input distribution that will be specified in the sequel.

 Recalling the discussion in Remark~\ref{remark1}, for the Gaussian RC with informed relay, we should consider two important features in the design of an efficient coding scheme at the relay: obtaining correlation or coherence between the channel inputs from the source and the relay, and exploiting the channel state to remove the effect of the CSI on the communication. As we already mentioned, it is not obvious to accomplish these features because the channel state is not available at the source. The main idea in the coding scheme that we consider consists in splitting the relay input $X_2^n$ into two independent parts, namely $U_1^n$ and $\tilde{X}_2^n$. The first part, $U_1^n$, is a function of only the cooperative information, and is generated using standard coding. Since the source knows the cooperative information at the relay, it can generate its codeword $X_1^n$ in such a way that it is coherent with $U_1^n$, i.e., by allowing correlation between $X_1^n$ and $U_1^n$. The second part, $\tilde{X}_2^n$, which is independent of the source input $X_1^n$, is a function of both the cooperative information and the channel state $S^n$ at the relay, and is generated using a GDPC similar to that in \cite{ZV07b, KL04, KL07a, SBSV07a, MS06}.

More formally, we decompose the relay input random variable $X_2$ as
\begin{equation}
X_2=U_1+\tilde{X}_2,
\label{RelayInputAchievableRateFullDuplexGaussianRC}
\end{equation}
where: $U_1$ is zero mean Gaussian with variance $\bar{\theta}P_2$, is independent of both $\tilde{X}_2$ and $S$, and is correlated with $X_1$ with $\mathbb{E}[U_1X_1]=\rho'_{12}\sqrt{\bar{\theta}P_1P_2}$, for some $\theta \in [0,1]$, $\rho'_{12} \in [-1,1]$ ; and $\tilde{X}_2$ is zero mean Gaussian with variance ${\theta}P_2$, is independent of $X_1$, and is correlated with the channel state $S$ with $\mathbb{E}[\tilde{X}_2S]=\rho'_{2s}\sqrt{{\theta}P_2Q}$, for some $\rho'_{2s} \in [-1,1]$.
Using the covariances $\sigma_{12}=\mathbb{E}[X_1X_2]=\mathbb{E}[X_1U_1]$ and $\sigma_{2s}=\mathbb{E}[X_2S]=\mathbb{E}[\tilde{X}_2S]$, the parameters $\rho'_{12}$, $\rho'_{2s}$ are given by
\begin{equation}
\rho'_{12}=\frac{\sigma_{12}}{\sqrt{\bar{\theta}P_1P_2}},\quad \rho'_{2s}=\frac{\sigma_{2s}}{\sqrt{{\theta}P_2Q}}.
\label{CorrelationCoefficientsFullDuplexGaussianRC}
\end{equation}
For the GDPC, we  choose the auxiliary random variable $U_2$ as
\begin{equation}
U_2=\tilde{X}_2+\alpha_{\text{opt}}S
\label{GeneralizedDPCatRelayFullDuplexGaussianRC}
\end{equation}
with
\begin{equation}
\alpha_{\text{opt}}=\frac{{\theta}P_2(1-\rho'^2_{2s})}{{\theta}P_2(1-\rho'^2_{2s})+N_3}\Big(1+\rho'_{2s}\sqrt{\frac{{\theta}P_2}{Q}}\Big)-\rho'_{2s}\sqrt{\frac{{\theta}P_2}{Q}}.
\label{OptimalCostaParameterGeneralizedDPCatRelayFullDuplexGaussianRC}
\end{equation}

We now provide an upper bound on the capacity of the state-dependent Gaussian RC with informed relay. 
\vspace{0.2cm}

\begin{theorem}\label{TheoremOuterBoundNonCausalCaseGaussianChannelFullDuplexRegime}
The capacity of the state-dependent general Gaussian RC with informed relay is upper bounded by
\begin{align}
& R^{\text{out}}_{\text{G}}=\max \min \Bigg\{\frac{1}{2}\log\Big(1+P_1(1-\frac{\rho^2_{12}}{1-\rho^2_{2s}})(\frac{1}{N_2}+\frac{1}{N_3})\Big),\nonumber\\
&\frac{1}{2}\log\Big(1+\frac{(\sqrt{P_1}+\rho_{12}\sqrt{P_2})^2}{P_2(1-\rho^2_{12}-\rho^2_{2s})+(\sqrt{Q}+\rho_{2s}\sqrt{P_2})^2+N_3}\Big)\nonumber\\
&\hspace{2.5cm}+\frac{1}{2}\log(1+\frac{P_2(1-\rho^2_{12}-\rho^2_{2s})}{N_3})\Bigg\},
\label{OuterBoundNonCausalCaseGaussianChannelFullDuplexRegime}
\end{align}
 where the maximization is over parameters $\rho_{12} \in [0,1]$ and $\rho_{2s} \in [-1,0]$ such that
\begin{equation}
\rho^2_{12}+\rho^2_{2s} \leq 1.
\label{AllowableCovarianceMatrixOuterBoundFullDuplex}
\end{equation}
\end{theorem}
\vspace{0.3cm}

The proof of Theorem \ref{TheoremOuterBoundNonCausalCaseGaussianChannelFullDuplexRegime} follows by evaluating the upper bound \eqref{OuterBoundNonCausalCaseDiscreteMemorylessChannel} using an appropriate joint distribution of $X_1,X_2,S,Y_2,Y_3$. It is based on showing that for the Gaussian channel \eqref{ChannelModelForFullDuplexRegimeGaussianRCWithState}, one can restrict attention to jointly Gaussian $(S,X_1,X_2,Y_2,Y_3)$ with $\mathbb{E}[X_1S]=0$, $\sigma_{12}=\rho_{12}\sqrt{P_1P_2}=\mathbb{E}[X_1X_2]$ and $\sigma_{2s}=\rho_{2s}\sqrt{P_2Q}=\mathbb{E}[X_2S]$. The allowable values for the covariances  $\sigma_{12}$ and  $\sigma_{2s}$ are such that the covariance matrix  $\Lambda_{X_1,X_2,S,Z_2,Z_3}$ of $(X_1,X_2,S,Z_2,Z_3)$ has a non-negative discriminant, i.e. $QP_1P_2N_2N_3(1-\rho^2_{12}-\rho^2_{2s}) \geq 0$. For $Q > 0$, this gives \eqref{AllowableCovarianceMatrixOuterBoundFullDuplex}. 

Similarly to in Theorem \ref{TheoremOuterBoundNonCausalCaseGaussianChannelFullDuplexRegime}, we obtain an upper bound on channel capacity for the degraded Gaussian case by evaluating the upper bound~\eqref{OuterBoundDegradedChannelNonCausalCaseDiscreteMemorylessChannel} in Corollary~\ref{CorollaryOuterBoundDegradedRelayChannelNonCausalCaseDiscreteMemorylessChannel} using an appropriate jointly Gaussian distribution of $S,X_1,X_2,Y_2,Y_3$.
\vspace{0.2cm}

\begin{corollary}\label{CorollaryUpperBoundGaussianChannelFullDuplexRegime}
The capacity of the state-dependent degraded Gaussian RC with informed relay is upper bounded by \eqref{OuterBoundNonCausalCaseGaussianChannelFullDuplexRegime} in which the first term of the minimization is replaced by
\begin{equation}
\frac{1}{2}\log\Big(1+\frac{P_1(1-\rho^2_{12}-\rho^2_{2s})}{N_2(1-\rho^2_{2s})}\Big).
\label{FirstTermUpperBoundDegradedGaussianChannelFullDuplexRegime}
\end{equation}
\end{corollary}
\vspace{0.3cm}

\subsection{Special Cases Analysis}\label{secV_subsecC}
We note that comparing the bounds in Theorem~\ref{TheoremAchievabeRateNonCausalCaseGaussianChannelFullDuplexRegime} and Theorem~\ref{TheoremOuterBoundNonCausalCaseGaussianChannelFullDuplexRegime} analytically can be tedious in the general case. In the following, we focus  only on the physically degraded case. In this case, we show that the lower bound in \eqref{AchievabeRateNonCausalCaseGaussianChannelFullDuplexRegime} is tight for  certain values of the channel statistics, and thus obtain the capacity expression for these cases.

In the following corollary we recast the upper bound  \eqref{OuterBoundNonCausalCaseGaussianChannelFullDuplexRegime} into an equivalent form by substituting $\kappa=\rho_{12}/\sqrt{1-\rho^2_{2s}}$ and $\rho=\rho_{2s}$. 

\begin{corollary}\label{CorollaryEquivalentFormOuterBoundNonCausalCaseDegradedGaussianChannelFullDuplexRegime}
For the state-dependent degraded Gaussian RC with informed relay, the upper bound \eqref{OuterBoundNonCausalCaseGaussianChannelFullDuplexRegime} can be written as
\begin{align}
& R^{\text{out}}_{\text{DG}} \:=\:\: \max_{\kappa} \min \Bigg\{\frac{1}{2}\log\Big(1+\frac{P_1(1-\kappa^2)}{N_2}\Big),\nonumber\\
&\max_{\rho} \frac{1}{2}\log(1+\frac{P_2(1-\kappa^2(1-\rho^2)-\rho^2)}{N_3})\nonumber\\
&+\frac{1}{2}\log\Big(1+\frac{P_1+\kappa^2(1-\rho^2)P_2+2\kappa\sqrt{1-\rho^2}\sqrt{P_1P_2}}{P_2(1-\kappa^2(1-\rho^2))+Q+2\rho\sqrt{P_2Q}+N_3}\Big)\Bigg\},
\label{EquivalentFormOuterBoundNonCausalCaseDegradedGaussianChannelFullDuplexRegime}
\end{align}
where the maximization is over parameters $\kappa \in [0,1]$ and $\rho \in [-1,0]$.\\
\end{corollary}
\vspace{-0.5cm}

Investigating the lower bound \eqref{AchievabeRateNonCausalCaseGaussianChannelFullDuplexRegime} and the upper bound \eqref{EquivalentFormOuterBoundNonCausalCaseDegradedGaussianChannelFullDuplexRegime}, it can be shown that the lower bound for the degraded case is tight for certain values of $P_1$, $P_2$, $Q$, $N_2$, $N_3$. The following proposition provides some cases for which the lower bound is tight.
 \begin{proposition}\label{CapacityForSpecialCases}
For the physically degraded Gaussian RC channel,  we have the following.\\
$1)$ If $P_1$, $P_2$, $Q$, $N_2$, $N_3$ satisfy
\begin{align}
N_2 \geq \max_{\zeta \in [-1,0]} \frac{P_1N_3(P_2+Q+N_3+2{\zeta}\sqrt{P_2Q})}{P_1N_3+P_2(1-\zeta^2)(P_1+P_2+Q+N_3+2{\zeta}\sqrt{P_2Q})},
\label{SnrRangeForChannelCapacityLowSnrAtRelay}
\end{align}
then channel capacity is given by
\begin{equation}
\mc C_{\text{DG}} = \frac{1}{2}\log(1+\frac{P_1}{N_2}),
\label{ChannelCapacityLowSnrAtRelay}
\end{equation}
which is the same as the interference-free capacity, i.e., the capacity when the channel state is also known to the source or is not present in the model.

$2)$ If the maximizing $\rho_{12}$ and $\rho_{2s}$ in the upper bound in Theorem  \ref{TheoremOuterBoundNonCausalCaseGaussianChannelFullDuplexRegime} are such that condition \eqref{AllowableCovarianceMatrixOuterBoundFullDuplex} is met with equality, i.e., $\rho^2_{12}+\rho^2_{2s}=1$, then the lower bound \eqref{AchievabeRateNonCausalCaseGaussianChannelFullDuplexRegime} is tight and gives capacity.
\end{proposition}
\noindent \textbf{Extreme Cases} \\
We now summarize the behavior of the capacity $\mc C_{\text{DG}}:=\mc C_{\text{DG}}(P_1,P_2,Q,N_2,N_3)$ in some extreme cases.
\begin{enumerate}
\item[1.] For $Q=0$, i.e., no channel state at all in the model, capacity is given by
\begin{align}
\mc C_{\text{DG}}=\max_{0\leq \beta \leq 1} \min \Big\{&\frac{1}{2}\log\Big(1+\frac{P_1(1-\beta^2)}{N_2}\Big),\nonumber\\
&\frac{1}{2}\log\Big(1+\frac{P_1+P_2+2\beta\sqrt{P_1P_2}}{N_3}\Big)\Big\},
\label{CapacityNoSideInformationDegradedGaussianChannelFullDuplexRegime}
\end{align}
which is the same as the capacity of the standard degraded Gaussian channel \cite[Theorem 5]{CG79}.
This can be directly obtained by substituting $Q=0$ in \eqref{AchievabeRateNonCausalCaseGaussianChannelFullDuplexRegime} and \eqref{EquivalentFormOuterBoundNonCausalCaseDegradedGaussianChannelFullDuplexRegime}. In this case, the maximizing parameters are $\theta=0$, $\rho'_{2s}=0$ for \eqref{AchievabeRateNonCausalCaseGaussianChannelFullDuplexRegime} and $\rho=0$ for \eqref{EquivalentFormOuterBoundNonCausalCaseDegradedGaussianChannelFullDuplexRegime}.

\item[2.] In the case of arbitrary strong channel state, i.e., $Q \rightarrow \infty$, capacity is given by
\begin{align}
& \mc C_{\text{DG}} \: =\:\: \min \Big\{\frac{1}{2}\log(1+\frac{P_1}{N_2}),\:\frac{1}{2}\log(1+\frac{P_2}{N_3})\Big\}.
\label{CapacityVeryStrongSideInformationDegradedGaussianChannelFullDuplexRegime}
\end{align}
In this case, the lower bound in \eqref{AchievabeRateNonCausalCaseGaussianChannelFullDuplexRegime} is maximized for $\theta=1,\rho'_{2s}=0,\rho'_{12}=0$, and the upper bound in \eqref{EquivalentFormOuterBoundNonCausalCaseDegradedGaussianChannelFullDuplexRegime} is maximized for $\rho=0,\kappa=0$, and the two bounds meet. We note that, in this strong channel state case, \eqref{CapacityVeryStrongSideInformationDegradedGaussianChannelFullDuplexRegime} suggests that traditional multi-hop transmission achieves the capacity. A multi-hop scheme allows to completely cancel the effect of the channel state by subtracting it out upon reception at the relay, and by applying standard DPC for transmission from the relay to the destination.
 
\item[3.] If $P_2=0$, capacity is given by
\begin{align}
& \mc C_{\text{DG}} \: =\:\: \frac{1}{2}\log(1+\frac{P_1}{Q+N_3}).
\label{CapacityZeroPowerAtRelayDegradedGaussianChannelFullDuplexRegime}
\end{align}
In this case, the informed relay cannot help the source, and the channel state is simply treated as an unknown noise.
\end{enumerate}
\vspace{-0.0cm}

\section{Numerical Examples and Concluding Remarks}\label{secVI}
In this section we discuss some numerical examples, for both the degraded Gaussian case and the general Gaussian case. 
\begin{figure}[!htpb]

        \begin{minipage}[t]{\linewidth}
        \vspace{-0.3cm}
        \begin{center}
        \includegraphics[width=\linewidth,height=0.6\linewidth]{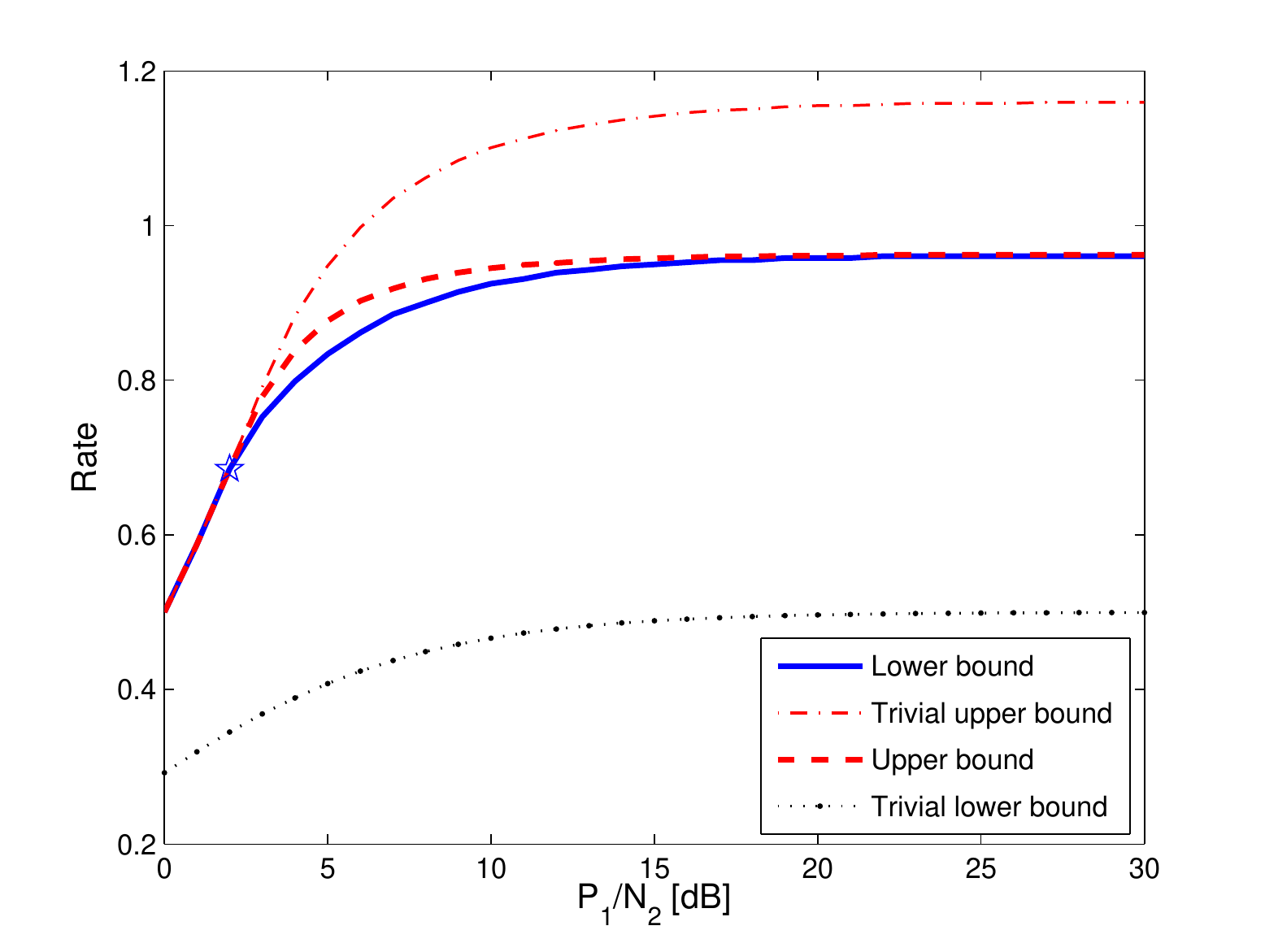}
        \label{Subfig2Fig1IllustrativeExamples}
        \vspace{-0.5cm}
        \caption{Lower and upper bounds on the capacity of the state-dependent degraded Gaussian RC with informed relay versus the SNR in the link source-to-relay. Numerical values are $P_1=P_2=Q=N_3=10$ dB. \vspace{-0.3cm}}
        \label{Fig1IllustrativeExamples}
        \end{center}
        \end{minipage}
\end{figure}

Figure~\ref{Fig1IllustrativeExamples} illustrates the lower bound \eqref{AchievabeRateNonCausalCaseGaussianChannelFullDuplexRegime} and the upper bound \eqref{EquivalentFormOuterBoundNonCausalCaseDegradedGaussianChannelFullDuplexRegime} as functions of the signal-to-noise-ratio (SNR) at the relay, i.e., $\text{SNR}=P_1/N_2$ (in decibels). Also shown for comparison are the trivial upper bound \eqref{TrivialOuterBoundNonCausalCaseDiscreteMemorylessChannel} computed for the degraded Gaussian case and the trivial lower bound obtained by considering the channel state as an unknown noise. The curves show that the lower bound and the upper bound do not meet for all SNR regimes. However, as it is visible from the depicted numerical examples, the gap between the two bounds is small for the degraded case. Furthermore, the curves in Figure~\ref{Fig1IllustrativeExamples} also illustrate the results in proposition \ref{CapacityForSpecialCases}, by showing that the lower bound and the upper bound meet for the cases stated in Proposition~\ref{CapacityForSpecialCases}. We note that the pentagram marker visible in Figure ~\ref{Fig1IllustrativeExamples} indicates capacity when the noise at the relay is equal to the RHS of \eqref{SnrRangeForChannelCapacityLowSnrAtRelay}.

\begin{figure}[!htpb]

        \begin{minipage}[t]{\linewidth}
        \vspace{-0.3cm}
        \begin{center}
        \includegraphics[width=\linewidth,height=0.8\linewidth]{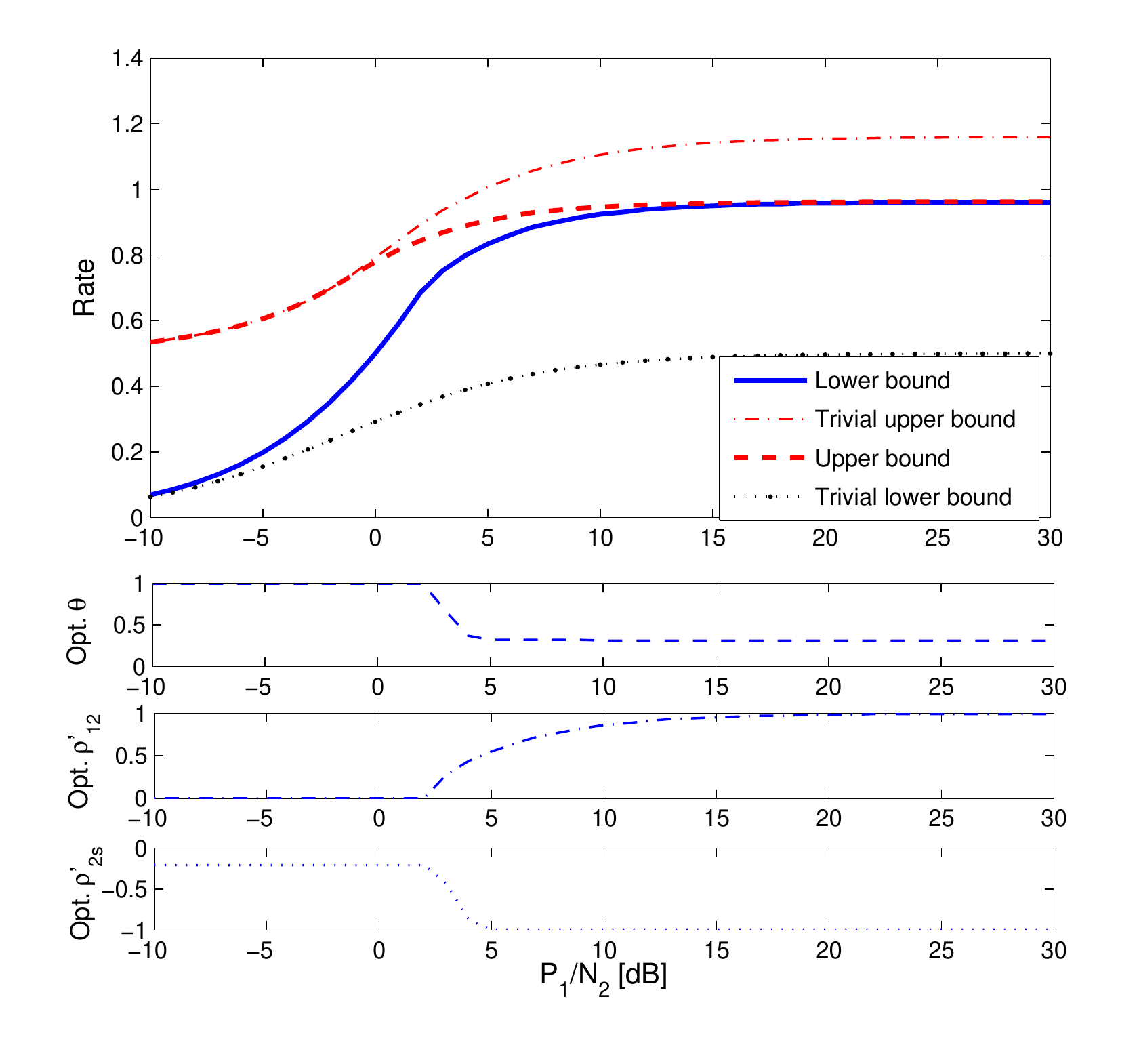}
        \vspace{-0.5cm}
       \caption{Lower and upper bounds on the capacity of the state-dependent general Gaussian RC with informed relay and the maximizing $\theta, \rho'_{12}, \rho'_{2s}$ in \eqref{AchievabeRateNonCausalCaseGaussianChannelFullDuplexRegime} as functions of the SNR at the relay. Numerical values are $P_1=P_2=Q=N_3=10$ dB.}
        \label{Fig5IllustrativeExamples}
        \end{center}
        \end{minipage}

\end{figure}
Figure~\ref{Fig5IllustrativeExamples} illustrates the lower bound \eqref{AchievabeRateNonCausalCaseGaussianChannelFullDuplexRegime} and the upper bound \eqref{OuterBoundNonCausalCaseGaussianChannelFullDuplexRegime} as functions of the $\text{SNR}$ at the relay for the general Gaussian channel. Also shown for comparison are the trivial upper bound \eqref{TrivialOuterBoundNonCausalCaseDiscreteMemorylessChannel} computed for the general Gaussian case and the trivial lower bound obtained by considering the channel state as an unknown noise. The curves show that the lower bound is close to the upper bound at large $\text{SNR}$, i.e., when capacity of the channel is determined by the sum rate of the MAC formed by transmission from the uninformed source and the informed relay to the destination. Furthermore, Figure~\ref{Fig5IllustrativeExamples} also shows the variation of the maximizing $\theta$, $\rho'_{12}$, $\rho'_{2s}$ in \eqref{AchievabeRateNonCausalCaseGaussianChannelFullDuplexRegime} as function of the $\text{SNR}$ at the relay. This shows how the informed relay allocates its power among combating the interference for the source (related to the value of $\rho'_{2s}$) and sending signals that are coherent with the transmission from the source (related to the values of $\theta$ and $\rho'_{12}$).

\vspace{-0.0cm}
\bibliography{paperITW2008}
\end{document}